# Nucleosome-mediated cooperativity between transcription factors.

Leonid A. Mirny

Harvard-MIT Division of Health Sciences and Technology, and Department of Physics, Massachusetts Institute of Technology, Cambridge, MA
E-mail: leonid@mit.edu Tel: ++1-617-452-4862

**Cooperative binding of transcription factors (TFs) to cis-regulatory regions (CRRs) is essential for precision of gene expression in development and other processes. The classical model of cooperativity requires direct interactions between TFs, thus constraining the arrangement of TFs sites in a CRR. On the contrary, genomic and functional studies demonstrate a great deal of flexibility in such arrangements with variable distances, numbers of sites, and identities of the involved TFs. Such flexibility is inconsistent with the cooperativity by direct interactions between TFs. Here we demonstrate that strong cooperativity among non-interacting TFs can be achieved by their competition with nucleosomes. We find that the mechanism of nucleosome-mediated cooperativity is mathematically identical to the Monod-Wyman-Changeux (MWC) model of cooperativity in hemoglobin. This surprising parallel provides deep insights, with parallels between heterotropic regulation of hemoglobin (e.g. Bohr effect) and roles of nucleosome-positioning sequences and chromatin modifications in gene regulation. Characterized mechanism is consistent with numerous experimental results, allows substantial flexibility in and modularity of CRRs, and provides a rationale for a broad range of genomic and evolutionary observations. Striking parallels between cooperativity in hemoglobin and in transcription regulation point at a new design principle that may be used in range of biological systems.**

## Introduction

In higher eukaryotes, CRRs are 200-1000 bps long and may contain clusters of 10-30 TF binding sites (TFBS)[1-3]. The arrangement, identity and affinity of the sites determine the function of the CRR. Cooperative binding of TFs to CRRs leads to highly cooperative gene activation and is essential for development [4] and other vital processes [5].

Cooperative binding is traditionally explained by protein-protein interactions among participating TFs [6, 7]. While this mechanism finds support in some bacterial and eukaryotic systems, several functional and genomic observations are inconsistent with it. The mechanism of cooperativity due to protein-protein interactions (directly or via DNA looping [6-8]) should significantly constrain arrangements of TFBS, allowing only those that provide the correct orientation, order and distance between TFs. On the contrary, recent evolutionary analysis of fly enhancers revealed massive rearrangements of TFBS [9] [10], and is further supported by experimental studies, which demonstrated that CRRs could tolerate incorporation of new binding sites (promiscuity) and significant alterations in TFBS placement while retaining *in vivo* functionality [9] [11, 12]. A few mechanisms proposed to explain flexible arrangements of TFBS and promiscuity are based on the idea of transcriptional synergy, i.e. cooperative recognition or simultaneously touching of TFs by some flexible part of the transcription machinery [11, 12], rather than on cooperative binding of TFs to DNA.

Here we present a mechanism of cooperativity among TFs mediated by nucleosomes. Nucleosomes pack DNA in eukaryotic cells, each consists of a histone protein core and 147 bps of DNA wrapped around it, and are believed to suppress gene activity. Presented mechanism demonstrates how cells can exploit nucleosomes to achieve cooperative DNA binding, while having flexible and promiscuous regulatory regions.

The phenomenon of nucleosome-mediated cooperativity have been documented by a series of *in vivo* and *in vitro* experiments [5, 13-15] which demonstrated synergistic binding and gene activation by non-endogenous TFs (e.g. Gal4 and LexA) that occupy sites on nucleosomal DNA (**Fig 1**). Such cooperativity requires only DNA-binding domain of TFs, suggesting that it does not involve chromatin modification or direct protein-protein interactions between TF [16]. Ploach and Widom proposed that binding of TFs to nucleosomal could provide to induced interactions between TFs: binding of the first TF leads to partial unwrapping of nucleosomal DNA thus making the site of the second TF more accessible [17]. Spontaneous partial unwrapping of nucleosomal DNA was later demonstrated biochemical in single-molecule and bulk experiments [18, 19]. While providing a proof of principle for nucleosome-induced cooperativity, the mechanism of assisted unwrapping works only as long as the DNA remains bound to a histone core [20] and is hardly applicable to CRRs that were shown to be mostly nucleosome-free in living cells.

Here we present a novel mechanism of nucleosome-mediated cooperativity and its quantitative model, which integrates binding on TF to nucleosomal and naked DNA, includes possible nucleosome displacement/eviction and establishing the role of TF is establishing nucleosome-free regulatory regions. We demonstrate that competition between nucleosome formation and an array of TF sites induces highly cooperative TF binding. Mathematically, the model is identical to the Monod-Wyman-Chaneux model of cooperativity in hemoglobin [21]. Using this striking analogy we gain deeper insights into a range of phenomena such as nucleosome positioning sequences and nucleosome modifications, the role of low-affinity TFBS and TFBS clustering, the origin of DNase hypersensitive sites (see **Table 1**). Next we show how presented mechanism allows significant flexibility of TFBS arrangements, easy substitution of new TFs, and CRR modularity. Finally we present several experimental evidences in support of this nucleosome-mediated mechanism.

## Results and Discussion

We consider interactions of TFs with a stable nucleosome (Fig. 1B), containing an array of $n$ TFBS located within its footprint (147bps) (**Fig.1**). This region of DNA can be in one of the two states: the nucleosome ($N$) and open ($O$) state when the histones are displaced or evicted from the region.

While histones limit access of other proteins to nucleosomal DNA, the nucleosome is highly dynamic with DNA unwrapping and wrapping back at very high rate, thus making nucleosomal DNA at least partially accessible to TFs [22, 23]. TFBS can be occupied in either state: $N_i$ and $O_i$ stand for the states with $i$ occupied sites. The affinity of TFs for the sites depends on the state with binding constants $K_N$ and $K_O$, respectively, and higher affinity in the open state: $c=K_O/K_N<<1$. For simplicity of the presentation we assume all TFBS to have the same affinity and experience the same suppression, *c*, by the nucleosome. These assumptions are dropped leading to more complicated equations presented in the supplement.

The equilibrium between *N* and *O* states in the absence of bound TFs is characterized by $L=[N_0]/[O_0]$, with $L>>1$ for a stable nucleosome. In equilibrium, the system is fully defined by the following three dimensionless parameters: *c, L,* and the aggregate concentration of TFs $\alpha=[TF]/K_O$ **(Figure 1B).** Based on experimental measurements we estimated them to be in the following range: $L=100\text{-}1000$, $c=10^{-1}\text{-}10^{-3}$ and $\alpha$=1-10 (see Supplement). For simplicity, we assume TFBS to be cognate sites of a single TF present in concentration [TF] (see Supplement for the general case).

Strikingly, the system of TFs and a nucleosome is identical to the scheme of cooperativity in hemoglobin as described by the classical MWC model [21]. **Table 1** summarizes equivalent parameters and analogous phenomena between the two systems. The MWC model considers equilibrium between two states of the hemoglobin tetramer: the R state, which has a higher affinity for $O_2$, and the low-affinity T state. In the absence of the oxygen, the hemoglobin is mostly in the T state. Binding of $O_2$ shifts the equilibrium toward the R state, making binding of successive oxygen molecules more likely and thus cooperative (**Fig 1C**). The R and T states of hemoglobin correspond to O and N states of the nucleosome, and the oxygen binding to individual hemoglobin domains corresponds to TFs binding individual sites (**Fig 1BC, Table 1**). This analogy helps us to gain deep understanding of cooperativity and regulation in the TF-nucleosome system.

The two quantities of primary biological interest are the occupancy per TFBS of the CRR by the TFs (*Y)* and the occupancy by the nucleosome (*Z)*. By analogy to MWC, and using tools of statistical mechanics we obtained expressions for these quantities:

$$Y=\alpha\frac{(1+\alpha)^{n-1}+Lc(1+c\alpha)^{n-1}}{(1+\alpha)^{n}+L(1+c\alpha)^{n}} \tag{1}$$

$$Z=\frac{L(1+c\alpha)^{n}}{(1+\alpha)^{n}+L(1+c\alpha)^{n}} \tag{2}$$

**Figure 2A** presents equilibrium TF and nucleosome occupancies *Y* and *Z* as a function of TF concentration, computed using experimentally determined parameters (see Supplementary Information for parameter estimation). Strikingly, nucleosome-mediated cooperativity can provide a sharp transition with a two-fold increase in TF concentration leading to a more than eight-fold increase in the occupancy. The nucleosome occupancy also changes cooperatively, dropping from about 65% to less than 10% due to a two-fold change in TF concentration.

Analogy to the MWC model allows us to reveal features of the nucleosome-TF system that are essential for cooperativity (**Table 1**). The MWC system has a strong cooperativity as long as *L* is sufficiently large ($L>10\text{-}100$) and *c* is sufficiently small ($c<0.1$). In our system (**Fig.2B**), this corresponds to requirements for nucleosome stability (i.e. forming a nucleosome more than 90% of the time, in the absence of TFs) and sufficient (at least ten-fold in $K_d$) attenuation of TF binding by a nucleosome. These requirements are consistent with high stability [24, 25] and slow exchange [26],[27] of nucleosomes in regions depleted in TFBS and unaffected by polymerase passage (e.g. inactive ORFs). *In vitro* studies of TF binding to nucleosomal DNA demonstrate the required attenuation of TF binding. **Figure 2B** shows that the cooperative transition is robust to variation in *L* and *c*, becoming, as expected from the MWC model, sharper at larger *L* and smaller *c*.

We also exploit analogy to MWC to examine the implications of sequence-specific nucleosome positioning, histone modifications and other processes involved in gene regulation. In fact, these effects can be considered as allosteric heterotropic regulation of the nucleosome-TF system, analogous to heterotropic effectors of hemoglobin. A prototypical heterotropic allosteric regulation in hemoglobin is the Bohr effect: lowering the pH causes the oxygen affinity to decrease, thus providing more oxygen to actively working muscles. The basis of the Bohr effect is in the higher affinity of hydrogen ions to the T state. Thus low pH stabilizes the T state, shifting the equilibrium away from the high-affinity R state. Other allosteric effectors of hemoglobin (e.g. DPG) act in a similar way: binding to one of the states of hemoglobin affects the R-T equilibrium and thus affinity of the hemoglobin to the oxygen.

Several processes in gene regulation are counterparts on the Bohr effect in hemoglobin: Histone modifications and histone-binding proteins affect nucleosome stability, thus altering the N-O equilibrium. To study this effect we used equations (1) and (2) to examine how the TF and nucleosomal occupancy curves change in response to reduced nucleosome stability *L.* **Figure 2C** shows manifestation of the “Bohr effect” in TF-nucleosome system: small changes in nucleosomal affinity (from *L* to *L’*) due to histone modifications can shift the balance in TF-nucleosome competition toward or away from the nucleosome. For example, a modification that reduces nucleosome stability by about $\Delta G=1$Kcal/mol ($\Delta G=k_B T\log(L/L')$) can lead to an 80% drop in nucleosome occupancy and a concurrent rise of the TF occupancy (**Fig. 2C,inset**). Nucleosome-positioning sequences have a similar effect: they alter nucleosome stability thus shifting the occupancy curve (**Fig 2C**) Importantly, cooperativity in nucleosome binding induced by competition with TFs leads to amplification of the sequence signal in CRRs, i.e. small changes in histone affinity translate into significant changes in nucleosome occupancy (**Fig. 2C,inset**). This effect helps to explain how weak nucleosome positioning sequence signals can lead to significant differences in nucleosome occupancy of different regions and sharp borders between the regions along the genome [26, 28].

Heterotropic regulation can also work in the opposite way: factors influencing binding of TFs can impact chromatin structure in TFBS-rich regions. For example, activation of a tissue-specific TF can lead to selective reduced chromatization and increased accessibility of tissue-specific CRRs (see **Fig.3**). This result is in agreement with a recent genome-scale mapping of DNaseI hypersensitive sites (DHS) that reported a highly tissue-specific DHS profile [29]. As discussed above, cooperativity of this allosteric regulation causes small changes in the concentrations of active TFs to significantly reduce nucleosomal occupancy (**Fig 2A and 3C**). Note that this mechanism of nucleosome displacement by competition with activated TFs does not rely on recruitment of chromatin modification machinery, which may play a role in further destabilizing nucleosomes [30] and expanding nucleosome-free regions initially formed by the competition. This effect will be considered in details elsewhere.

Nucleosome-induced cooperativity, however, has some properties that do not have counterparts in hemoglobin. For example, the number and the strength of biding sites is constant in hemoglobin, but vary in CRRs. **Figure 2D** presents nucleosomal occupancy as a function on the number of TFBS calculated using equations (1) and (2). As the number of TFBS exceeds a certain critical value $n_c$, nucleosomal occupancy drops sharply, manifesting another allosteric effect in the system. Our calculations show that the critical number of TFBS is given by $n_c \approx \log(L)/\log(1+\alpha)$, yielding a narrow range $n_c$=3-6 that is not very sensitive to model parameters (see Supplement). **Figure 3** demonstrates that clusters of 5 high-affinity and 8 low-affinity TFBS become occupied and nucleosome-free, while isolated sites remain TF-free. Having the number of sites in this range per nucleosomal footprint is both sufficient for cooperativity and consistent with the recent genomic characterization of *Drosophila* CRRs [1, 3] that contain several clusters of 4-6 sites in a region of 100-150 bps [2]**.** This effect can also explain widespread depletion and rapid exchange of nucleosomes in TFBS-rich CRRs [24, 26, 31-33].

Our approach allows generalization of MWC cooperativity to consider contributions of low-affinity sites, mixtures of sites of different TFs etc (see Supplement for equations and derivations). For example, assuming that only a few TFs are sufficient for activation, as was demonstrated experimentally [34], we obtain the expression for the probability of having at least $k$=2 of $n$ sites being occupied, $P_k$ (**Fig 3**). We also considered the contribution of low-affinity sites that were shown to play an important role in fly enhancers [35, 36]. **Figure 3** illustrates how arrays of low-affinity and high-affinity sites in nucleosomal DNA respond differently to increasing level of TFs. Our formalism allow one to calculate experimentally observable nucleosomal and TF occupancy of a DNA region with the number and strength of sites that can be altered experimentally.

Below we discuss several experimental results that support nucleosome-mediated mechanism of cooperativity, summarized in **Table 2**. Most direct evidences come for experimental studies that demonstrated cooperative binding of TFs without involvement of direct protein-protein interactions for a range of up to 200 bps [14]. Moreover, experiments with TFs lacking activation domains have shown that synergetic activation of gene expression is determined more by the number of TFBS than the interactions with general TFs, polymerase or a machinery of chromatin modification [34, 37]. Consistent with the nucleosome-induced mechanism, trans-complementation experiments on stripe 2 enhancers showed that precise expression does not require special Bcd-Hb interactions and can be achieved by chimeric and non-endogenous (i.e. non-interacting) TFs [36]. The range (~150-200 bps) of nucleosome-induced cooperativity is also consistent with the reported modularity of the *otd* enhancer, which contains two 180 bp TFBS clusters, each able to provide the correct expression pattern [38].

The presented mechanism also helps to tie together several observations in functional and comparative genomics. Cooperative nucleosome displacement can explain how low nucleosomal density is maintained on CRRs and promoters and how sharp boundaries of such nucleosome-depleted regions are achieved. The critical number of sites $n_c$, calculated above, required for the TF-induced nucleosome displacement is consistent with clustered arrangements of TFBS and helps to explain why such clustering serves as a powerful criterion for bioinformatic identification of CRRs [39]. The nucleosome-mediated mechanism suggests a role for low-affinity TFBS, which are essential in fly enhancers [36] and abundant [2, 40], in assisting high-affinity sites to displace nucleosomes and provide cooperative binding. This mechanism serves, along with sequence information [41, 42], in providing nucleosome positioning, but in contrast, can be tissue/condition specific, as recently reported [29].

The nucleosome-mediated mechanism provides significant evolutionary flexibility to CRRs, allowing considerable rearrangements of TFBS while retaining cooperativity. Such widespread flexibility and rapid evolution of CRRs has been reported [1-3] and was hard to reconcile with the classical model of cooperativity by direct protein-protein interactions. Moreover, the described mechanism explains observed promiscuity of CRRs: it allows unrelated TFs to cooperate in gene regulation, simply by evolving TFBS close to each other in CRR. Similarly, TFs can become a part of an existing assembly by acquiring TFBS within an existing cluster, a fairly fast and widespread evolutionary process [10, 43]. A classical model of cooperativity would also require interacting TFs to evolve interfaces that facilitate protein-protein interactions – a process that requires many more mutations.

Competition with a single nucleosome can provide cooperative binding for TFBS separated at most 150-200bps. This range can be further increased by the demonstrated synergy of nearby nucleosomes [44] and can spread much further through recruitment of chromatin modification machinery and due to positive feedback in this process [45].

In summary, we have showed how competition between a nucleosome and TFs can lead to cooperative binding of TFs and cooperative displacement of nucleosomes from regulatory regions. We have established and employed the analogy between this process and cooperativity in hemoglobin according to MWC model. This analogy has allowed us to consider chromatin modification and nucleosome positioning sequences as heterotropic allosteric effectors, similar to the Bohr effect. Most importantly, the presented mechanism explains a wealth of genomic and evolutionary observations that cannot be reconciled with the classical model of cooperativity among TFs. Our study provides strong support to the view that regulatory regions are flexible and highly evolvable regions of the genome. Finally, the analogy between cooperativity in hemoglobin and nucleosome-mediated cooperativity of TFs hints at the possible universality of the MWC mechanism of cooperativity in seemingly unrelated biological processes.

## Acknowledgments

I am grateful to, Meharn Kardar, Stanislav Shvartsman, Paul Wiggins, Jane Kondev, Ned Wingreen, and particularly to George Benedek for insightful discussions of this study, and to Jason Leith and Zeba Wunderlich for discussing and editing the manuscript.

# Figure 1

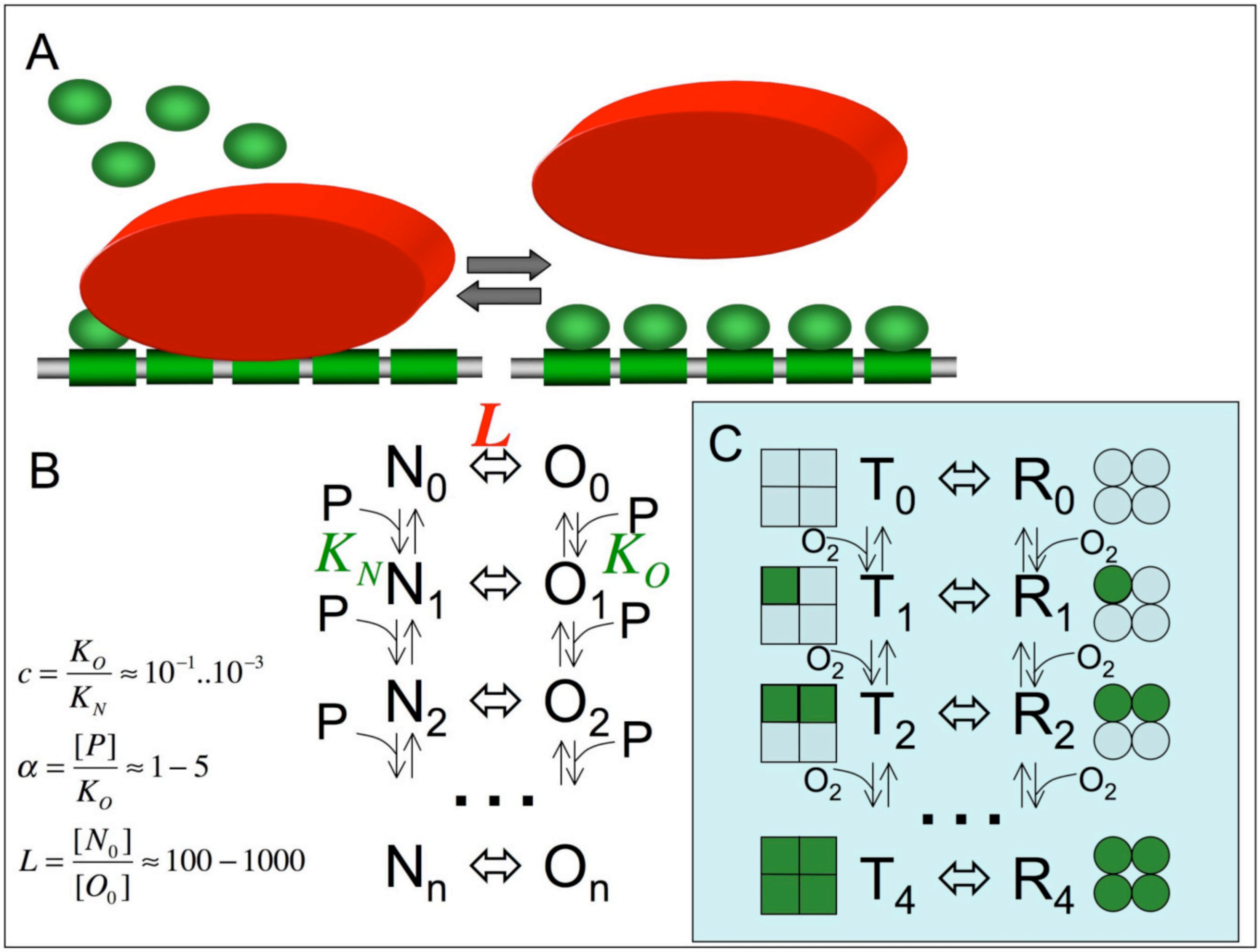

**Figure 1. The model of nucleosome-mediated cooperativity. A.** A region of DNA that contains an array of ***n*** sites (green boxes) can be bound by a histone core (red oval), thus becoming a nucleosomal DNA, or remain naked. In either nucleosomal (N) or open (O) state, the DNA can be bound by transcription factors (TFs, green ovals). Binding of TFs to the nucleosomal DNA is diminished as compared to the naked DNA ($K_O >> K_N$), but is possible due to transient, partial unwinding of the DNA (and not nucleosome translocation) [46]. The system is in thermal equilibrium and is fully characterized by the scheme in **B.** Each state of the system is determined by the form of the DNA: nucleosomal ($N_i$) and open ($O_i$), with ***i*** being the number of TFs bound. In this form, the nucleosome-TF system is identical to the Monod-Wyman-Changeux (MWC) model of cooperativity in hemoglobin (**C**). The N and O forms of the DNA correspond to T and R states of the hemoglobin; binding of TFs is equivalent to binding of the oxygen molecule; the attenuation of TF affinity to their sites in the nucleosome corresponds to weaker affinity for the oxygen in the T state. At a low concentration of TFs (low oxygen pressure), the DNA is mostly in the nucleosomal state (the hemoglobin is in the T state). At a high concentration of TFs (oxygen), binding of ligand pulls the equilibrium toward the open form (R state), displacing the nucleosome. The key feature of the MWC model is the coordinated transition between R and T states of all four domains, corresponding to binding of the histone octamer to the whole stretch of 147bps of DNA. Like MWC, the nucleosome-TF system is determined by three dimensionless constants: *L*, *c* and $\alpha$ that respectively control the preference of the N (T) state in the absence of the ligand, attenuated affinity for the ligand in this state, and the ligand concentration normalized to affinity for a site in the O (R) state. The range of values is estimated using experimentally measured quantities (see Supplementary Information).

# Figure 2

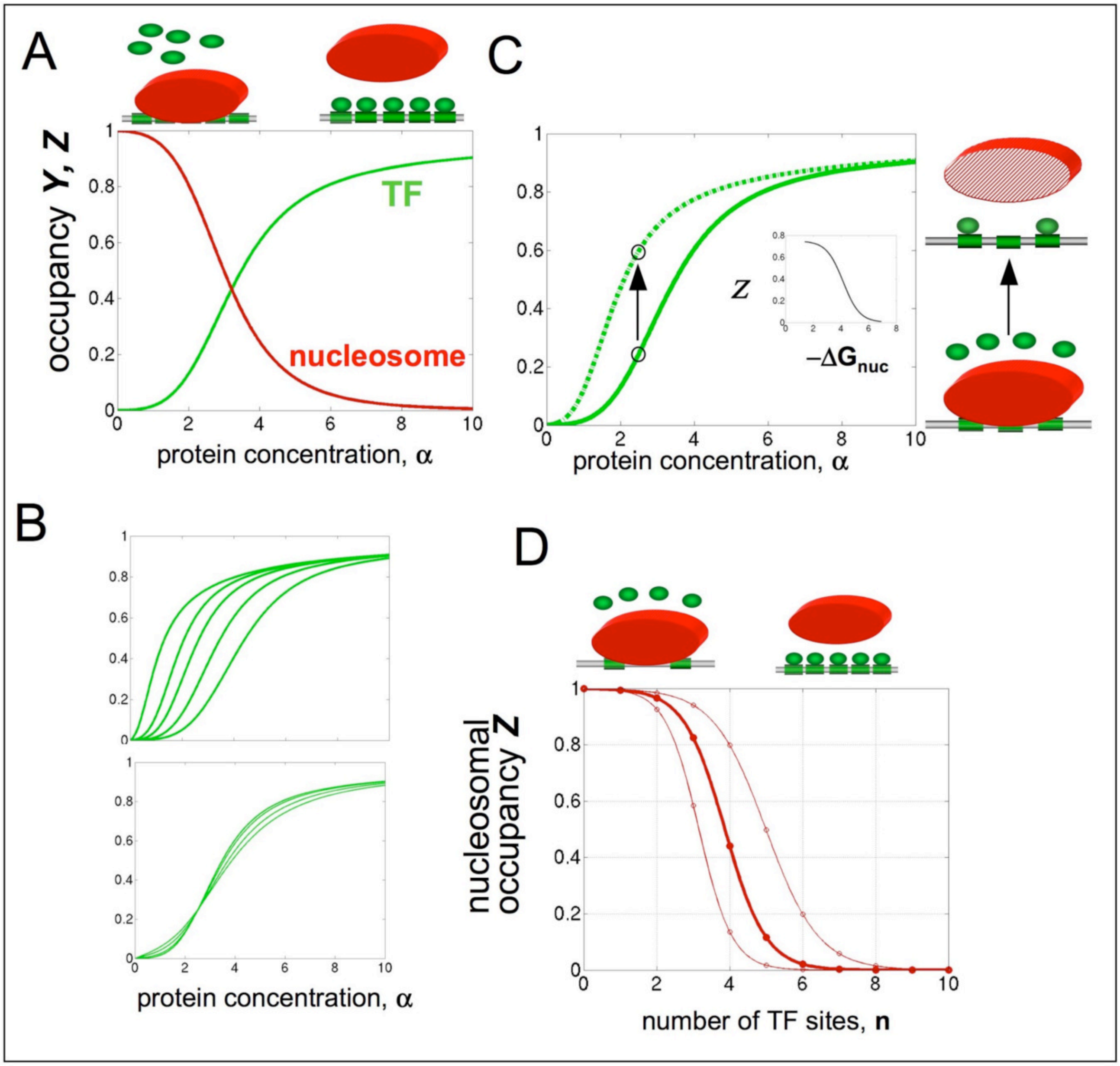

**Figure 2. Nucleosome-mediated cooperativity and its implications. A.** Cooperative transition in the equilibrium occupancy of TFs, *Y* (green line) and nucleosome *Z* (red line) as function of TF concentration (eq.1 and 2). Here and below $n=5$, $L=10^3$, $c=10^{-3}$, unless stated otherwise. **B.** Robustness of the cooperativity to 300-fold variation in parameters *L (top, L=10-3000)* and c *(bottom, c=0.03-10^-4)*. **C.** Bohr effect: attenuation of nucleosomal core affinity for DNA, due to modifications or as a function of DNA sequence, leads to a shift of balance in TF-nucleosome competition and displacement of the nucleosome by TFs (arrow). This competition makes nucleosomal occupancy, Z, respond considerably to small changes in nucleosome affinity **(inset),** as demonstrated by the dependence of Z on $-\Delta G=k_B T\log(L)$ *(Kcal/mol)* (see text for details). **D.** Effect of the number of TF sites, ***n***, on nucleosome stability, obtained for three concentrations of TF: $\alpha=3,5,8$. There is a clear critical number of sites (~4-5) below which TFs are unable to displace a nucleosome and above which the nucleosome in unstable even at a low concentration of TFs. This effect explains the clustering of sites in regulatory regions and demonstrates the origin of chromatin hypersensitive sites among such regions.

# Figure 3

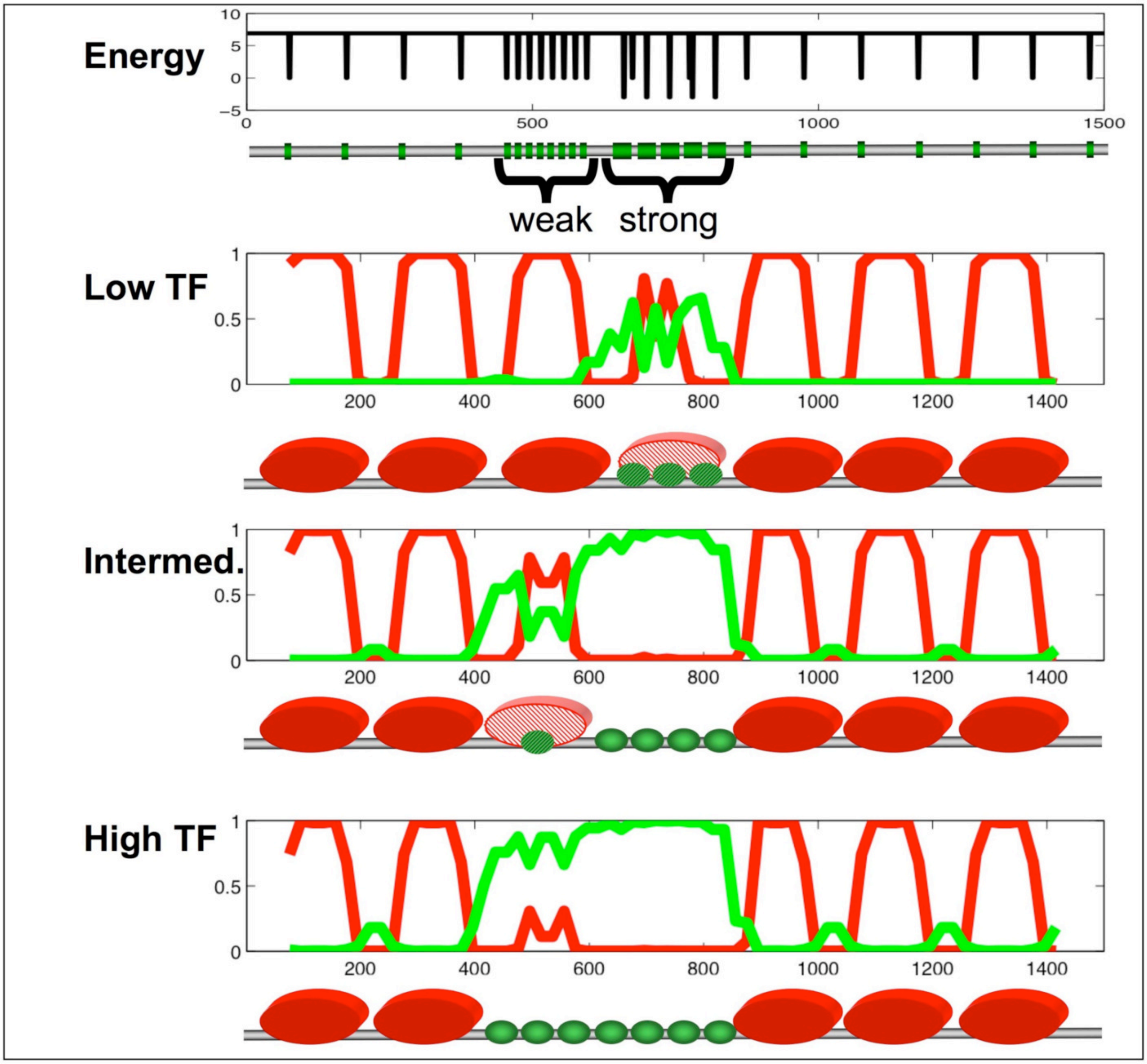

**Figure 3. Cooperative binding to high- and low-affinity sites.** The nucleosomal (red) and TF (green) occupancy profiles at a CRR that contains high-affinity (strong) and low-affinity (weak) sites. The top diagram shows the binding energy profile: a cluster of 8 low-affinity sites and a cluster of 5 high-affinity sites located over the background of scattered low-affinity sites. The CRR is packed by nucleosomes (nucleosomal binding profile is not shown). Three diagrams show nucleosomal occupancy *Z* (red), and TF cluster occupancy $P_3$ (probability of having at least three sites in a cluster occupied, green) for three values of TF concentration (see Supplement). These profiles show that at TFBS rich clusters can become nucleosome-free at the intermediate concentrations of TFs (see text for comparison to experiments). While intermediate TF concentration is sufficient to get high-affinity clusters nucleosome-free and TF-bound, a higher concentration may be needed for clusters of low-affinity site. Combination of low- and high-affinity sites in a CRR can make it respond differently to different TF concentrations. Notice that a cluster of several low-affinity TF is required to destabilize a nucleosome; an isolated low-affinity site is unable to do this.

# Tables

**Table 1. Comparison of the nucleosome-mediated cooperativity and cooperative transition in hemoglobin.**

| **Cooperative transition in hemoglobin, Monod-Wyman-Changuex model** | **Nucleosome-mediated cooperativity of transcription factors** |
|---|---|
| ***Components of the system*** | |
| Oxygen pressure, $pO_2$ | Concentration of TF, *[P]* |
| Two states of hemoglobin monomer: high affinity **R** and low affinity **T** states | Two states of DNA: high affinity **O** and low affinity **N** (open or nucleosome) states |
| Prevalence of the **T** state at low $pO_2$ (L>>1) | Prevalence of the **N** state (stable nucleosome) at low TF concentration (L>>1) |
| Four oxygen-binding hemes (n=4) | n TF binding sites |
| Allosteric transition in hemoglobin | Nucleosome assembly and displacement |
| ***Phenomena*** | |
| Cooperative binding of the oxygen | Cooperative binding of TF |
| Bohr effect | Concentration of histones and their affinity for DNA |
| Other heterotropic regulation | Histone modifications, sequence-dependent nucleosome stability, histone-binding proteins |
| Homotropic regulation | TF-dependent nucleosome depletion, interaction between different TFs |
| Energy stored In protein/heme deformation | Energy stored in DNA deformation and histone-DNA interactions |

**Table 2.** Comparison of the model's predictions with experiments.

| Model of the nucleosome-mediated cooperativity | Experimental data |
|---|---|
| Cooperative binding by non-interacting TFs | Gal4, USF and NF-kappa B bind cooperatively to reconstituted nucleosome *in vitro*. |
| Cooperative action of TF that bind within a footprint of a nucleosome <150bps, and independent of site orientations. | Cooperativity of Gal4 and USF independent of site orientations [15]. LexA and Gal4, with one TF lacking the activation domain, cooperate up to a range of 200bps. |
| Lack of cooperative binding in the absence of the nucleosomes | NF-kappa B act synergistically at a promoter containing four site. Binding is not cooperative in vitro [5] |
| Cooperative binding of TFs does not require direct interactions between them. | Structure of interferon-β enhanceosome demonstrating very few direct contacts between bound TFs [47]. |
| Displacement of a nucleosome occluding TFBS-rich regulatory region | Enrichment of TFBS in nucleosome depleted regions [29, 48], nucleosome depletion in yeast regulatory regions (promoters) [24, 26]; |
| Critical number ($n_c$=4-6) of TF binding sites required for cooperative binding. | Clustering of TF binding sites in drosophila enhancers, exciding 20 sites per Kb [2, 3]. Importance of clustering in eve2 [36]. |
| High concentration of TFs is required to displace a stable nucleosome, lower for modified nucleosome (Bohr effect) | Recruitment of chromatin remodeling is required for activation through low-affinity sites. Overexpression compensates for the lack of remodeling [49, 50]. Overexpression of TF compensates for mutation in high-affinity site, leading to nucleosome eviction through binding to two low-affinity sites [51]. |
| Nucleosomal occupancy depends on the presence of TFBS | Nucleosomal occupancy is restored by mutations eliminating TFBS [48]. Mutation in the high-affinity site of HSF reduces nucleosome eviction in vivo [51]. |

# Supplemental Information
# A. Derivation of equations

## Derivation of the TF and nucleosomal occupancy: MWC model

Here we use a statistical mechanics approach to derive occupancy and other equilibrium properties of the system. Alternative derivations can be found elsewhere [?]. The advantage of our approach is that it allows direct generalization for sites of different strength.

Consider $n$ sites, a protein having a concentration $P$, and an affinity to the site characterized by the binding constant $K$. Since the sites are bound independently, the probability of each site being occupied is

$$y = \frac{P}{P+K} = \frac{\alpha}{1+\alpha},$$

where $\alpha = P/K$ is a dimensionless protein concentration. It is easy to see that $\alpha$ is simply a statistical weight of the bound state.

The set of sites can be in two states, N and O, that determine the binding constants of all the sites: $K_N$ and $K_O$. The statistical weights of the bound site in each state are: $\alpha_O = P/K_O \equiv \alpha$ and $\alpha_N = P/K_N = cP/K_O = \alpha c$, where $c = K_O/K_N$ is another dimensionless parameters of the system.

The system in has $2n$ states: $N_0, N_1, ...N_n, O_0, O_1, ...O_n$, where the subscript stands for the number of occupied sites. The states $N$ and $O$ have different energies, and in the absence of any sites occuped the concentrations of the two states are connected by $L = N_0/O_0$ .

Thus the system is fully defined by three dimensionless parameters: $\alpha$, $c$, and $L$. First we calculate the equilibrium occupancy *per site*, i.e.

$$Y = \frac{1}{n}\frac{\sum_{i=1}^{n} i[w(O_i) + w(N_i)]}{Z} \tag{1}$$

with the partition function $Z = \sum_{i=1}^{n}[w(O_i) + w(N_i)]$, where $w(\cdot)$ is a statistical weight of each state:

$$w(O_i) = C_n^i \alpha_O^i = C_n^i \alpha^i \tag{2}$$
$$w(N_i) = LC_n^i \alpha_N^i = LC_n^i (\alpha c)^i, \tag{3}$$

where $C_n^k$ is the binomial coefficient and $L$ takes care of the higher statistical weight of the $N$ state.

Sums in the numerator and denominator can be easily calculated giving a closed form solution

$$Y = \alpha \frac{(1+\alpha)^{n-1} + Lc\,(1+c\alpha)^{n-1}}{(1+\alpha)^n + L\,(1+c\alpha)^n} \tag{4}$$

Eq **??** is identical to the mean occupancy of hemoglobin sites obtained in the MWC model.

We also calculate quantities that were not obtained in MWC, such as the nucleosomal occupancy as

$$Y_N = \frac{\sum_{i=1}^n w(N_i)}{Z} = \frac{L\,(1+c\alpha)^n}{(1+\alpha)^n + L\,(1+c\alpha)^n} \tag{5}$$

and the probability to have exactly $k$ sites occupied

$$p_k = \frac{kC_n^k \alpha^k \left(1 + Lc^k\right)}{(1+\alpha)^n + L\,(1+c\alpha)^n} \tag{6}$$

or at least $k$ sites occupied:

$$P_k = \sum_{i=k}^n p_i = \frac{\sum_{i=k}^n C_n^i \alpha^i \left(1 + Lc^i\right)}{(1+\alpha)^n + L\,(1+c\alpha)^n} \tag{7}$$

These quantities are particularly useful for dealing with large clusters of sites, where a few bound TF can be sufficient to activate transcription. For example, the probability of having at least one site occupied in a cluster is

$$P_1 = \frac{(1+\alpha)^n - 1 + L\left[(1+\alpha c)^n - 1\right]}{(1+\alpha)^n + Lc(1+\alpha)^n} = 1 - \frac{1+L}{(1+\alpha)^n + Lc(1+\alpha)^n}.$$

## Derivation of the TF and nucleosomal occupancy for distinct sites: generalization of MWC model.

Model presented above can be easily generalized for the case when TFBS have different strength, or when two or more types of TFs are poised to bind their respective TFBS in the region of interest. These cases lead to different statistical weights of different TFBS, i.e. $\alpha_i, i = 1..n$. We we were unable to obtain closed form solutions for such case, but obtained the following equations that can be treated numerically:

$$
\begin{aligned}
Y &= \frac{1}{n}\frac{\sum_{k=1}^{n}\alpha_k\prod_{i=1,i\neq k}^{n}(1+\alpha_i)+Lc\sum_{k=1}^{n}\alpha_k\prod_{i=1,i\neq k}^{n}(1+c\alpha_i)}{Z}, && (8)\\
Y_N &= \frac{L\prod_{i=1}^{n}(1+c\alpha_i)}{Z}, && (9)\\
Z &= \prod_{i=1}^{n}(1+\alpha_i)+L\prod_{i=1}^{n}(1+c\alpha_i)\,. && (10)\\
& && (11)
\end{aligned}
$$

Calculating $P_k$ becomes more problematic, but equations for practically important $P_1$, $P_2$, and $P_3$ can be obtained:

$$
\begin{aligned}
P_1 &= 1-\frac{1+L}{Z}, && (12)\\
P_2 &= 1-\frac{1+L+(1+Lc)\sum_{i=1}^{n}\alpha_i}{Z}. && (13)
\end{aligned}
$$

## Derivation of equation for the critical number of TFBS $n_c$ in a cluster.

Here we study how the CRR occupancy depends on the number of sites $n$. We focus on the nucleosomal occupancy, looking to find $n_c$ a critical value of sites sufficient to displace a nucleosome. We seek $n_c$ that provides $Y_N(n_c) = 0.5$. We consider a case where $c \ll 1$ and $\alpha \approx 0 - 10$ and thus can approximate $Y_N$:

$$
\begin{aligned}
Y_N &= \frac{L(1+c\alpha)^n}{(1+\alpha)^n+L(1+c\alpha)^n} \approx \frac{L}{L+(1+\alpha)^n}\\
Y_N(n_c) &= \frac{1}{2}\\
(1+\alpha)^{n_c} &= L\\
n_c &= \frac{\log L}{\log(1+\alpha)}.
\end{aligned}
$$

Using a range of values $L = 100 - 1000$ and $\alpha \approx 2 - 5$, we obtain $n_c \approx 3 - 6$. Note that this corresponds to $3 - 6$ TFBS *per* $\sim 200$ *bps*, i.e. $7 - 15$ TFBS for a CRR of 500bps, required to displace nucleosomes from this region.

# Supplemental Information B.

# B. Estimation of parameters using experimentally measured quantities.

### 1. Kd and TF concentration.

Here we estimate an effective concentration of TFs $\alpha$=[P]/K that enters all our equation. For a simple binding reaction between a cognate site S and a TF (a protein) P, the occupancy of the site is

$$P + S \rightleftarrows PS; \quad K_D = \frac{[P][S]}{[PS]}$$

$$Y = \frac{PS}{PS+S} = \frac{1/K_D}{1/K_D + 1/P} = \frac{P/K_D}{P/K_D + 1} = \frac{\alpha}{1+\alpha}$$

This quantity however does not take into account the effect of non-specific binding to all other DNA in the nucleus. Such non-specific binding is very important as it sequesters TFs from binding to the cognate site(s). Below we consider both specific and non-specific binding and show that later can be take into account by introducing an effective dissociation constant $K_D^{eff}$ to replace $K_D$.

Consider a TF (a protein) present at concentration [P] in the volume where non-specific DNA is present in concentration [DNA] and the cognate site(s) are present t the concentration [S]. The following binding reactions take place:

$$P + DNA \rightleftarrows P{\bullet}DNA$$

$$P + S \rightleftarrows PS$$

By solving:

$$K_D = \frac{[P][S]}{[PS]}; \quad K_D^{NS} = \frac{[P][DNA]}{[P{\bullet}DNA]}; \quad Y = \frac{[PS]}{[PS]+[S]}$$

we obtain

$$Y = \frac{1}{1 + DNA/K_D^{NS} \cdot K_D/P} = \frac{1}{1 + K_D^{eff}/P}$$

where

$$K_D^{eff} = K_D \cdot [DNA] / K_D^{NS}$$

Thus we parameters alpha should be defined using $K_D^{eff}$, and then the occupancy of the cognate site in the presence of non-specific DNA is as above:

$$\boxed{Y = \frac{\alpha}{1+\alpha}, \qquad \alpha = [P]/K_D^{eff} = [P]K_D^{NS}/[DNA]K_D}.$$

The dissociation constant of most eukaryotic TFs is in the range of $K_D \approx 1-10\text{nm}$, while available non-specific binding constants are about 1000 fold greater $K_D \approx 1-10\mu\text{m}$

Estimating alpha for yeast and tissue culture cells we have

***<u>Yeast</u>***
[P]≈500-2000 proteins per nucleus {cite YGFP, BiologyNumbers}
$K_D$≈1-10 nm; $K_D$≈1-10 um
[DNA]=12x$10^6$ bps per nucleus x 10% accessibility due to chromatization ≈ 1x$10^6$ bps per nucleus

An alternative way to estimate [DNA] is to multiply the length of nucleosome-free promoter regions measured experimentally to be approximately 200-300bps by the number of genes (6000) yielding = 1-2x$10^6$ bps

α=500-2000 x1000 / 1-2e6 **= 0.5-2**

***<u>Mammalian cells</u>***
[P]≈10,000-100,000 proteins per nucleus {cite BiologyNumbers, p53 conc}
[DNA]=2-3e9 bps x 10% accessibility = 0.2-0.3e9 bps

Alternatively: 30,000 genes x 1Kbps DHS (DNase hypersensitive) = 0.03e9 bps

α=1e4-1e5 x1000 / 200-30e6 **= 0.05-3**

Protein concentrations and binding constants used above are rather approximate and may be different for different TF and depend on the level of TF activation, its localization etc. Assuming that the protein concentration can go up upon activation and/or re-localization into the nucleus we considered a range **α=1-10**.

**2. Nucleosome stability, L.**
One way to estimate L is to use nucleosome equilibrium occupancy $f$.

$$f = \frac{[N]}{[N]+[O]}, \quad L = \frac{[N]}{[O]} = \frac{f}{1-f}$$

For stable nucleosomes occupancy is very close to 1 and hard to measure. We used two sources for $f$:

(1) Using *in vivo* measured fraction of bound H2B >98% as a proxy for f>0.98 **{Phair}**, we obtained L>50.
(2) Computed occupancy used for evaluation of nucleosome positioning by Sigal et al **{Sigal}** and Morozov et al **{Morozov}**. Using data from their Nucleosome Explorer site http://edsc.rockefeller.edu/nucleosome/ we obtained the equilibrium occupancy of stable nucleosomes f=0.98-0.999, yielding L=50-1000.

Low exchange rate and long residence time of H2B and stable nucleosomes *in vivo (residence time > 1 hour)* is consistent with such high stability of uperturbed nucleosomes.

Although imprecise these estimates of L dare sufficient to provide cooperativity. MWC formalism is very robust to changes in L, requiring L>>1 and Lc>1 for the onset of cooperativity.

**3. Suppression of TF binding and site exposure in a nucleosome, parameter c.**

Parameter *c* of our model reflects suppression of TF binding by an assembled nucleosome. Such suppression acts through steric hindrance between histones and TFs. The hindrance, however, is not permanent and TF sites on the nucleosomal DNA get exposed for TF binding. This mechanism was quantitatively characterized by *in vitro* experiments of Widom et al {Widom_review2008}. These experiments also demonstrated that such exposure of nucleosomal DNA does not require nucleosome disassembly, thus acting in the *N* state of our model. Thus suppression of binding in the N state, parameter c, is equivalent to experimentally measured equilibrium constant of site exposure. This constant depends on the location of the site with respect to the center of the nucleosome and has the following range:

$$c = 2\times10^{-2} - 10^{-5}$$

with $2\times10^{-2}$ for sites just inside from an end to $10^{-4}$-$10^{-5}$ for sites located near nucleosomal dyad. The lower the value of *c,* the stronger is the cooperativity of TF binding. In our study we have chosen one conservative value of ***c=0.01*** for all sites. Having different values of *c* for different sites is a straightforward generalization.

# Parameters used for Fig.3

c=0.01

L=1000 (for nucleosome positions) and 1e-7 (otherwise)

$\alpha_{non\text{-}site}$= 0.001;

$\alpha_{high\text{-}affinity}$ = 20;

$\alpha_{low\text{-}affinity}$= 1.